\documentclass[aps,superscriptaddress,pra]{revtex4-1}   
\usepackage{times,amsmath,amssymb,amsfonts,graphicx,color}
\usepackage{graphicx}   
\usepackage{dcolumn}    
\usepackage{bm}         
\usepackage{amssymb} 
\usepackage{color}

\newcommand{\ket}[1]{\vert #1 \rangle} %

\begin{document}
\title{Quantum correlations and synchronization measures}

\author{Fernando Galve}
\affiliation{IFISC (UIB-CSIC), Instituto de
F\'{\i}sica Interdisciplinar y Sistemas Complejos,  Palma de Mallorca, Spain}

\author{Gian Luca Giorgi}
\affiliation{IFISC (UIB-CSIC), Instituto de
F\'{\i}sica Interdisciplinar y Sistemas Complejos,  Palma de Mallorca, Spain}

\author{Roberta Zambrini}
\affiliation{IFISC (UIB-CSIC), Instituto de
F\'{\i}sica Interdisciplinar y Sistemas Complejos, Palma de Mallorca, Spain}

\date{\today}

\begin{abstract}

The phenomenon of spontaneous synchronization is universal and
only recently advances have been made in the quantum domain. Being synchronization a kind
of temporal correlation among systems, it is interesting to understand its connection
with other measures of quantum correlations.
We review here what is known in the field, putting emphasis on measures and indicators
of synchronization which have been proposed in the literature, and comparing their validity
for different dynamical systems, highlighting when they give similar insights and when they seem
to fail.

\end{abstract}

\maketitle

\tableofcontents

\section{Introduction}

Synchronization phenomena \cite{Pikovsky2001,Strogatz2001} and quantum 
correlations \cite{Horodecki2009,Modi2012,Adesso2016} have been studied for a long time by two 
different communities, and only recently their relation started to be 
explored. The common ingredient for the emergence of both features
is the mutual interaction between the components of a system, and in the 
quantum regime the potential relation between 
synchronization and the presence of mutual information, discord, entanglement 
or other correlations has been recently explored. 

The phenomenon of spontaneous or mutual synchronization refers to the ability  
of two or more systems, that would display different dynamics when separate, to 
evolve coherently when coupled. In the case of oscillatory dynamics this 
corresponds to achieving oscillation at a common frequency. This concept 
has been further refined and generalized  in chaotic systems to encompass several 
scenarios such as, for instance,  lag synchronization, generalized 
synchronization, or phase synchronization \cite{Boccaletti2002}. In 
general,  the definition of classical synchronization  itself refers to some  
similarity in the time evolutions, i.e. some temporal $correlation$ between the local
dynamical variables of the involved  systems. Therefore, this is a definition 
associated to classical trajectories. The counterpart, and eventually
generalization, in the quantum regime can follow different approaches. 

The first works on the subject of quantum synchronization were actually dealing 
with entrainment,  where an external driving acts as a pacemaker, in systems 
such as spin-boson with modulated driving \cite{Goychuk2006}, 
driven resonator with one \cite{Zhirov2008}  or two superconducting 
qubits \cite{Zhirov2009}, and more recently driven quantum Van der Pol 
oscillators  \cite{Lee2013a,Walter2014}. In the case of entrainment, or 
forced synchronization, the driver is a strong external field, generally 
classical, and is not influenced by the interaction with the system. Different is the 
case of mutual synchronization that refers to the emergence of synchronization 
as a collective phenomenon, leading to a coherent dynamics out of different 
coupled units, in the absence of a driver. 

Mutual synchronization has been recently predicted for spins interacting with a 
common bath \cite{Orth2010} and for the average positions of quantum 
optomechanical systems \cite{Heinrich2011,Holmes2012}.
The first analysis in the quantum regime dealt with
harmonic networks \cite{Giorgi2012,Manzano2013} looking at quantum noise 
synchronization 
and showing its counterpart in relation with classical and quantum correlations,  
 showing the same trend for mutual information and quantum discord. 
Synchronization in the 
 dynamics of second-order quadratures (squeezing)  was considered there with an exact approach and  
role of local, global and independent baths was elucidated.
In Ref.\cite{Mari2013} it was instead considered the question about the limits to perfect 
synchronization imposed by quantum fluctuations and uncertainty relations. 
The phenomenon  is characterized with a synchronization error and is discussed in the context of 
coupled optomechanical devices. 
Also in Ref.\cite{Lee2013a} there is a discussion of coherently coupled Van der Pol 
oscillators characterizing synchronization through their phase-locking in phase space. 
 
When extending the concepts of synchronization into the quantum regime, a first 
question is about what defines this phenomenon, as actually in the classical 
regimes it refers to the dynamics of
phase space trajectories and, in general, classical variables.
This chapter reviews this question showing different approaches as well 
as the specific peculiarities reported so far for quantum synchronization with 
respect to the classical case. 
In the following section we give a brief overview of the platforms where 
quantum 
synchronization is under 
study and then review the characterizations and measures proposed for this 
phenomenon.
We then discuss some general questions and possible future directions.

 \section{Synchronization in quantum systems}\label{sec2}

\begin{figure}[b!]
\includegraphics[width=0.9\textwidth]{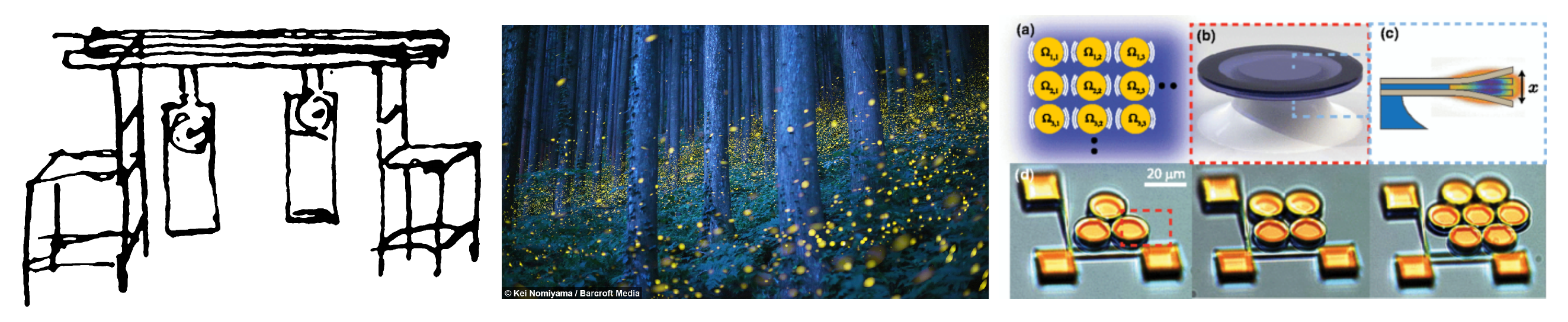}
\caption{A historical perspective of synchronization. 
Left panel: Original drawing of Huygens of two synchronizing pendulum 
clocks attached to a common support. Middle panel: 
Swarms of fireflies illuminate the undergrowth in a 
forest (photo by Kei Nomiyama/ Barcroft Media). 
Right panel: Micromechanical oscillator arrays coupled through light (figure taken from Ref. \cite{Zhang2015}).  }
\label{FigIntro}
\end{figure}

The phenomenon of synchronization is paradigmatic in a large variety of 
biological, physical, and chemical systems, operating in the classical regime,
as reviewed for instance in 
\cite{Pikovsky2001,Strogatz2001,Boccaletti2002,Manrubia,Arenas2008}. 
Some fascinating examples
are fireflies flashing at once (Fig.1), cardiac myocytes acting as pacemakers, or
the swaying motion of the millennium bridge due to the crowd walk in synchrony. 
The first reported observation of classical synchronization was described as 
``sympathy of two clocks'' and 
dates back to XVII century, when Huygens observed pendula hanging on a wall in 
a boat
(see extracts and references in \cite{Pikovsky2001}).
A reproduction of Huygens' original drawing is presented in Fig.\ref{FigIntro}.
An equivalent popular experiment is with metronomes on the same bar 
placed on two cylinders (cans) free to roll \cite{Pantaleone2002}.
The extension to chaotic systems has also been a wide field of research, 
establishing the possibility to observe this phenomenon in spite of the high 
sensitivity  to small differences in the initial conditions or 
device 
parameters \cite{Boccaletti2002}.

When moving to microscopic systems, synchronization phenomena are expected to 
take place, and 
several recent works address quantum synchronization in nanomechanical devices,
harmonic oscillators and spin systems.

\subsection{Nanomechanical devices}

Optomechanical devices exploit radiation pressure to couple  coherently light 
and matter motional degrees of freedom allowing to explore different aspects of 
synchronization in a flexible, hybrid and highly sensitive platform, where 
operation in the quantum regime has been achieved \cite{Aspelmeyer2014}. 
Spontaneous synchronization of optomechanical devices has been predicted 
theoretically considering mechanical coupling \cite{Heinrich2011} as well as 
coupling through a common optical mode \cite{Holmes2012}, focusing on the 
average positions
and  laying the base for the study of quantum signatures of synchronization in 
these devices. 
Phase-coherent mechanical oscillations have been shown in regular optomechanical 
crystals
considering the effects of quantum noise \cite{Ludwig2013}.

Reported experiments with microdisks 
\cite{Zhang2012} and arrays \cite{Zhang2015} and nanomechanical resonators 
interacting through
an optical racetrack \cite{Bagheri2013} display 
synchronization of the average (classical) motional degree of freedom. 
In Fig. \ref{FigIntro} the device used in Ref.  \cite{Zhang2015} is reproduced.
The  possibility to lock distant optomechanical oscillators has also been 
explored in the classical regime \cite{Shah2015,Li2016}. 
Recently it was also reported the 
experimental realization of spontaneous synchronization among micro- 
\cite{Agrawal2013}
and nano-electromechanical \cite{Matheny2014} autonomous oscillators.
Quantum signatures of synchronization phenomena in experiments on 
optomechanical devices have not yet been 
reported.

\subsection{Linear and non-linear oscillators}

Among theoretical models, both linear and non-linear oscillators have been 
considered theoretically in the quantum regime. 
Van der Pol oscillators  have been investigated in the quantum regime 
\cite{Lee2013a,Lee2014,Walter2014,Walter2015} and in 
comparison with the classical one \cite{Pikovsky2001}.
These models exhibit self-sustained oscillations and spontaneous
synchronization due to coherent \cite{Lee2013a,Ameri2015} and dissipative 
coupling 
\cite{Lee2014,Walter2015}, as well as 
phase locking  \cite{Lee2013a} and frequency entrainment  \cite{Walter2014} in 
presence of external drive.
The realization of Van der Pol oscillators  in physical platforms operating in 
the quantum regime has been suggested in  trapped ions \cite{Lee2013a} and 
optomechanical oscillators \cite{Walter2015}.  

Self-sustained oscillators, like Van der Pol oscillators are a well-known 
platform for studying synchronization. However, due to their non-linearity the 
analysis in the quantum regime can be only addressed in limited cases and under
various approximations, such as truncation of the Wigner function,
linearization around stable states \cite{Carmichael2009} or in the limit of 
infinite 
non-linear couplings favoring few low-energy Fock states
\cite{Lee2013a}. An $exact$ analysis can be performed in linear systems
like harmonic networks; these have been considered in order to 
identify the conditions for quantum synchronization beyond approximations and to 
clarify the role of 
losses, comparing diffusive and reactive couplings in Refs.
\cite{Giorgi2012,Manzano2013,Manzano2013a}.
The analysis of networks in squeezed vacuum \cite{Manzano2013} shows
that under dissipation in separate equivalent baths (a common assumption), 
independently on the strength of the 
coupling the oscillators will not be able to synchronize, while in any other 
more complex 
dissipation scenarios (common bath, local bath, etc...), the presence of one 
less 
damped normal mode of the system allows for  a transient or asymptotic 
synchronization. By accessing few oscillators parameters
this synchronization in the squeeezing dynamics can be tuned in the network or 
in clusters.

\subsection{Spin models}
As mentioned before, quantum spin synchronization was first 
discussed under the perspective of entrainment to an external driving force, 
either in the general  spin-boson framework  \cite{Goychuk2006} or considering 
superconducting two-level systems \cite{Zhirov2008,Zhirov2009}. 
An experimental observation was recently 
reported considering a damped current-biased Josephson junction \cite{Xue2014}.

Studies of spontaneous synchronization of spins within abstract theoretical 
models were performed in Refs. \cite{Orth2010,Giorgi2013} considering 
spin-boson dissipation. In Ref.  \cite{Orth2010}, it was first observed that
synchronization is induced by the coherent exchange of bath excitations between
the two spins, while in Ref. \cite{Giorgi2013} it was shown that  pure dephasing
is not able to generate synchronization. The formation of Chimera states was discussed in  \cite{Viennot2016} considering 
an extended spin chain described by a non-Hermitian Hamiltonian. 
In Ref. \cite{Ameri2015}, the authors  analyzed 
the behavior of two qubits placed inside two coupled  cavities where only the 
first one is driven by a laser. 
The steady-state synchronization of ensembles of dissipative, driven two-level 
atoms collectively coupled to a cavity mode, was studied in \cite{Xu2014} and, 
under more general conditions in 
\cite{Zhu2015}, where the authors also provided a direct analogy with the 
synchronization of classical phase oscillators.
Following the experimental results of Ref. \cite{Deutsch2010}, where a 
self-rephasing mechanism was observed on the ground state of magnetically 
trapped ultracold atoms, synchronization within a full quantum model for the 
case of two non-dissipative, interacting macro-spins, was studied in Ref.
\cite{Liu2013}.

Finally, a platform to probe synchronization was introduced in Ref. 
\cite{Hush2015}. There, the authors considered two cold ions in microtraps and 
studied the synchronization between their motional degrees of freedom.
The presence of synchronization was witnessed by the correlations developed by 
the electronic, discrete, degrees of freedom of the two ions. 
 
\subsection{Applications}

Synchronization  is clearly a resource in biological systems 
\cite{Pikovsky2001,Strogatz2001}.
The synchronized flashing of
fireflies is a strategy so that the female can identify her species-specific 
flashing signal (Fig. \ref{FigIntro}). 
Synchronization of neuronal activity by phase locking of self-generated network 
oscillations dynamics 
is one of the coordinating mechanisms of the brain, and abnormalities in this 
process are at the basis of several diseases and dysfunctions,
like epilepsy. 

The achievement of a coherent dynamics out of different components (also due to 
experimental imperfections) is clearly a resource also in physical systems. An  
interesting application, for instance, 
is in cryptographic protocols based on chaotic carriers of signals  
\cite{Argyris2005}.
In general, synchronization allows for enhancing of frequency stability, 
coherence and power output. Therefore applications are envisaged 
for precise frequency sources, time-keeping, and sensing 
\cite{Matheny2014,Zhang2015} and can be taken also to the domain of quantum 
technologies. 

An application of a synchronization transition was recently proposed
as  an effective tool to 
probe the dynamics of a quantum system dissipating through a thermal bath 
\cite{Giorgi2016}.  Indeed, coupling the system to an external, detuned object 
(which plays the role of the probe)  a transition between in-phase and 
anti-phase system-probe  synchronization is observed as a function of the 
detuning and of the spectral density of the bath. Clearly, this transition can 
be observed monitoring  the dynamics of the probe alone. Then, measuring the 
critical detuning at which the transition takes place amounts to getting 
information about the whole dissipative process.

\section{Measures of mutual quantum synchronization}\label{sec3}

Synchronization refers to some coherence in the temporal dynamics of coupled
systems and several measures are known in the classical realm \cite{Pikovsky2001,Strogatz2001,Boccaletti2002}.
Synchronization in presence of driving, entrainment, is typically encoded in the
phase
locking of the slave system with respect to the drive: the detuning between the slave oscillation
and the driving frequency is
typically
plotted as a function of the frequency of the driver to identify the region of
locking (zero detuning) and when this region's width is considered for different
driving strengths one gets the synchronization region known as Arnold tongue
\cite{Pikovsky2001}.

In autonomous systems, synchronization can arise as a mutual phenomenon, the
final dynamics coming from the interaction between components.
The equivalent to an Arnold tongue appears by considering the relative coupling and
detuning of the system components.
Despite its intuitive conceptual definition, the quantification of
synchronization in the quantum realm is a challenging
problem where both temporal and quantum correlations come into play.
Two or more objects, irrespective of their quantum or classical
nature, do spontaneously
synchronize if they adjust their own local dynamics to a common pace
determined by their mutual
interaction. Then, a good synchronization measure is expected to be able to
capture this adjustment of rhythms, that can only
be detected monitoring the behavior of all local units. Synchronization can be
inferred observing how similar
the local density matrices are, according to some meaningful criteria, or
considering local observables and looking at
their correlations in time. Furthermore, it is sometimes possible to deduce the
behavior of local observables inspecting
overall quantities, like, for instance, emission spectra.

A broad plethora of studies of synchronization in classical systems in the last three decades
provides useful hints about possible approaches when moving into the quantum
regime. The main difference with respect to classical systems is clearly that
synchronization there generally refers to time trajectories and
limit cycles in the phase space not present in the quantum approach. On the other hand, classical synchronization
has been already generalized in presence
of noise and of chaos \cite{Boccaletti2002} where it is identified by assessing the
'similarity' of local dynamical evolutions quantified by several indicators.
Indeed a number of these indicators can be
taken into the quantum regime providing insightful approaches to quantum
synchronization,
as it is the case of the Pearson function and synchronization error introduced
below.

These considerations do not exclude that manifestations of synchronization can
be found also in global indicators, including mutual information and
correlations, and that can be associated to genuine quantum properties of the
whole state.
Overall, when addressing the question of the identification of genuine quantum
synchronization phenomena, two main approaches can be distinguished: one is to
define synchronization in local observables and look for the presence of
quantum correlations triggered by this phenomenon; an alternative approach is
to define synchronization itself as a form of quantum correlation. In the following we
give an overview of different physical cases where synchronization is expected to
come out, discussing the interplay between local indicators and collective ones.

\subsection{Pearson factor}
\label{Pearson}
\label{Sec_Pearson}

The Pearson's  correlation coefficient is a  widely used  measure of the degree
of linear dependence between two variables. Calling $X$ and $Y$ the variables,
it is defined as
\begin{equation}
{\cal C}_{X,Y}=\frac{E[XY]-E[X]E[Y]}{\sqrt{E[X^2]-E[X]^2}\sqrt{E[Y^2]-E[Y]^2}},
\end{equation}
where $E[.]$ is an average value. As a consequence of the definition,
${\cal C}_{X,Y}$ gives a value between +1 and −1, where 1 indicates total
positive correlation, 0 is the absence of correlation, and −1 is total negative
correlation. Considering two functions $f(t)$ and $g(t)$ evolving in time, the
Pearson's  coefficient ${\cal C}_{f(t),g(t)}(t)$ can be calculated over a
sliding window of length $ \Delta t$ replacing the expectation values with time
averages: in this case
\begin{equation}
E[f](t,\Delta t)\equiv \overline{f(t,\Delta t)}=\frac{1}{\Delta
t}\int_t^{t+\Delta t} f(t).
\end{equation}
 Given two time-dependent variables $A_1$ and $A_2$ the Pearson synchronization measure reads
\begin{equation}\label{eq:pears}
 {\cal C}_{A_1,A_2}(t|\Delta t)=\frac{\int_{t}^{t+\Delta t}(A_1-\bar{A_1})(A_2-\bar{A_2})dt}
 {\sqrt{ \int_{t}^{t+\Delta t}(A_1-\bar{A_1})^2 dt  { \int_{t}^{t+\Delta t}(A_2-\bar{A_2})^2 dt}}},
\end{equation} 
where
\begin{equation}
 \bar{A_i}=\frac{1}{\Delta t}\int_{t}^{t+\Delta t}A_idt.
\end{equation} 
By definition, this measure quantifies the temporal correlation between two classical
trajectories and has been widely used in classical synchronization problems
\cite{Pikovsky2001,Boccaletti2002}. 

In the quantum framework the trajectories $A_i$ can be the expectation values of quantum operators, as moments at different orders  of 
local observables, like
$\langle \hat N_i\rangle,\langle \hat x_i^2\rangle, \langle \hat x_i^4\rangle$, $\sigma_i^x$....
This measure was first adopted in the framework of quantum synchronization in Ref.
\cite{Giorgi2012}, using the quantum-mechanical expectation values of the
second moments of the positions and momenta of two linearly coupled harmonic
oscillators  dissipating in a common environment. In this way the
synchronization in the dynamics of second-order quadratures  was captured. The same quantification of
synchronization was also carried out in the case of an extended network of
linear harmonic oscillators  \cite{Manzano2013}.

In the case of
dissipating spin pairs, the Pearson's coefficient was adopted  in Refs.
\cite{Giorgi2013,Giorgi2016} to quantify the degree of synchronization between
$\langle \sigma_1^x\rangle$ and $\langle \sigma_2^x\rangle$. A slightly
different version of it, especially tailored to detect phase synchronization,
was also analyzed in Ref. \cite{Ameri2015}. A qualitative analysis of the
similarity between the time evolution of local averages of spin operators, even
though without any explicit reference to the Pearson's measure,  was also
invoked by Orth {\textit{et al.}} \cite{Orth2010}.  

In general, this measure can be applied to any quantum problem when looking at temporal dynamics of
local observables. The main advantages are (i) that it depends on the quantum signatures of the system (e.g.
quantum noise, when going beyond first order moments) and (ii) that this measure has absolute reference
values: reaching the maximum (minimum) value
${\cal C}_{X,Y}=1 (-1)$ for perfect (anti-)synchronization.

\subsection{Synchronization error} \label{Sec_error}

The averaged distance between classical trajectories has been largely used to study synchronization of chaotic systems, see e.g.
the example of a pair of bidirectionally coupled Lorenz
systems in Ref. \cite{Boccaletti2002} of two coupled cahotic systems. This synchronization error
was first considered in \cite{Mari2013} to study quantum synchronization of coupled optomechanical oscillators
attaining limit cycles. For continuous variable (CV) systems the synchronization error reads
\begin{equation}
\mathcal{S}_c(t)=\left< q_-(t)^2+p_-(t)^2\right>^{-1},\label{syncerr}
\end{equation}
where $q_-=(q_1-q_2)/\sqrt{2}$ is the difference in position, and the same for momentum, of the objects of interest. 
At variance with the classical case where the average distance can go to zero, in the quantum domain this measure is bounded.
It achieves
a maximum value when the two quantum objects are synchronized,  and is upper-bounded by the uncertainty principle
\begin{equation}\label{upper}
\mathcal{S}_c(t)\leq\frac{1}{2\sqrt{\left< q_-(t)^2\right>\left<p_-(t)^2\right>}}\leq 1.
\end{equation}
A poor value of this quantity can come from two possible origins: either the mean value (first moment) of $q_-$ and $p_-$ is big, or because the variances of these
operators are big. In order to neglect the first cause, it is also interesting to define a modified measure 
with
\begin{equation}
 q_-(t)\rightarrow q_-(t)-\left< q_-(t)\right>\, ,\,  p_-(t)\rightarrow p_-(t)-\left< p_-(t)\right>,
\end{equation}
which is preferable if we want to study purely quantum effects.

While synchronization error in classical systems is generaly addressed between the time series of two deterministic variables, 
as for example $q_1(t)$ and $q_2(t)$, the instrinsic probabilistic nature in the quantum domain enlarge this scenario.
As for the Pearson factor, when comparing two operators $\hat{q}_1(t)$ and $\hat{q}_2(t)$ (hats are omitted elsewhere), the corresponsing
first moments can behave independently of the second moments or moments of higher order. This is particularly the case for Brownian oscillators initialized in 
vacuum squeezed states.
The intention of the authors in \cite{Mari2013} is to be able to compare the two operators by introducing a quadratic error measure, as reported in
optomechanical settings (see \ref{Sect_Self-sustainedOM} section), gauging well (as can be seen from comparison to other synchronization indicators) both the synchronization of 
first moments and second moments. The relation of this measure with the synchronization of local dynamics is still an open question.

In a similar spirit a measure of phase synchronization is also introduced in \cite{Mari2013}, by writing the operator $a_j(t):=[q_j(t)+ip_j(t)]/\sqrt {2}$ of the $j$th system in the following way
\begin{equation}
a_j(t))=[r_j(t)+a'_j(t)]e^{i\phi_j(t)},
\end{equation}
where $r_j$ and $\phi_j$ are the amplitude and phase of the expectation value of $a_j(t)$: $\left<a_j(t)\right> = r_j(t)e^{i\phi_j(t)}$. Now the hermitian and anti-hermitian parts of
$a'_j(t):=[q'_j(t)+ip'_j(t)]/\sqrt {2}$ can be interpreted as amplitude and phase fluctuations, and we can say that whenever $\left<a_1(t)\right>$ and $\left<a_2(t)\right>$ are phase 
locked we can define a phase shift with respect to this locking by the operator $p'_-(t)=[p'_1(t)-p'_2(t)]/\sqrt{2}$. Hence, a measure of phase synchronization is
\begin{equation}
\mathcal{S}_p(t)=\frac{1}{2}\left<p'_-(t)^2\right>^{-1},
\end{equation}
which in contrast to $\mathcal{S}_c$ can be arbitrarily large \cite{Mari2013}. The authors point out though that $\mathcal{S}_p\leq 1$ whenever two CV quantum systems can be represented by a positive
$P$ function (quantum optics notion of classicality), whereas the opposite would require collective squeezing.

\subsection{Mutual information and other information-based correlations}

Entropic measures are often used in different contexts to  quantify the 
correlation between sub-parts. In many cases, these quantities have been 
compared to other classical synchronization measures. Here we briefly review the case 
where they have been proposed in relation to quantum synchronization. 

Classical mutual information, associated to time series of system observables,
has been used as a measure of classical synchronization \cite{Boccaletti2002}.
The quantum mutual information ($MI$) of a whole density matrix $\rho_{AB}$ is 
defined as 
  \begin{equation} 
I(\rho)=S(\rho_A)+S(\rho_B)-S(\rho_{AB}),
\end{equation} 
where $\rho_A$ ($\rho_B$) is the reduced density matrix obtained tracing the 
subpart $B$ ($A$) and where $S$ stands for the von Neumann entropy: 
$S(\rho)=-{\rm Tr}\{\rho \log \rho\}$. $MI$ was proposed in Ref. 
\cite{Ameri2015} as synchronization witness. The authors considered two 
different models showing synchronization, that is, two coupled Van der Pol 
oscillators and two qubits inside optical cavities in the presence of driving. 
It was shown that the steady-state $MI$  had the same qualitative
behavior of, respectively, the  complete synchronization measure of Eq. (\ref{syncerr}), and   
relative phase between the two qubits (measured with an indicator close to 
Pearson's parameter) during  the transient. Comparisons between mutual 
information and synchronization had already been performed in 
Refs. \cite{Giorgi2012,Manzano2013,Giorgi2013} showing that in harmonic systems  \cite{Giorgi2012,Manzano2013} 
$MI$ is more robust in the synchronization regime, while for 
spins coupled through the environment \cite{Giorgi2013} it is not distinctive signature of synchronization. 

In Refs. \cite{Giorgi2012,Manzano2013,Giorgi2013,Ameri2015}, synchronization 
was also compared to quantum discord, the part of mutual information 
quantifying nonclassical correlations \cite{Ollivier2001,Henderson2001},
leading to similar results. 
Given a bipartite system $AB$, it is defined as the difference between 
$I(\rho)$ and the classical part of correlations 
${\cal{J}}(\rho)_{\{\Pi_j^B\}}=S(\rho_A)-S(A|\{\Pi_j^B\})$,
where the conditional entropy is 
$S(A|\{\Pi_j^B\})=\sum_ip_iS(\varrho_{A|\Pi_i^B})$, 
$p_i={\rm Tr}_{AB}(\Pi_i^B\varrho)$ and where 
$\varrho_{A|\Pi_i^B}= \Pi_i^B\varrho\Pi_i^B/{p_i} $ 
is the density
matrix after an optimal, complete projective measurement 
$(\{\Pi_j^B\})$ has been performed on B.

Generalized versions of $MI$ can be obtained using  the R\'enyi entropy 
$S_\alpha(\rho)=(1-\alpha)^{-1}\log{\rm Tr}\{\rho^\alpha\}$ (which reduces to 
$S$ in the limit of $\alpha\to 1$). The R\'enyi-2 mutual information 
($I_2(\rho)=S_2(\rho_A)+S_2(\rho_B)-S_2(\rho_{AB})$) was used by Bastidas 
{\textit at al.} to detect chimera-type synchronization in a quantum network  
of  coupled Van der Pol oscillators \cite{Bastidas2015}. 
Chimera states describe the coexistence of synchronize and unsynchronized components
\cite{Motter2010}.

Entanglement has also been considered in the context of synchronization. 
Lee and coworkers, studying the case of two dissipatively coupled Van der Pol 
oscillators, argued that the steady-state exhibits an entanglement tongue, 
the quantum analogue of the Arnold tongue \cite{Lee2014}. 
Entanglement, after a truncation of the total Hilbert space of the 
two oscillators, was quantified using the concurrence $E$ \cite{Wootters1998}.
The concurrence between a pair of qubits, whose density matrix is $\rho$, is 
defined 
as $E={\rm max\lbrace 0,\lambda_1-\lambda_2-\lambda_3-\lambda_4\rbrace}$, where 
$\lambda_r$ is the square root of the $r$th eigenvalue of $R=\rho\tilde{\rho}$ 
in descending order. Here, we have introduced 
$\tilde{\rho}=(\sigma_y\otimes\sigma_y)\rho^*(\sigma_y\otimes\sigma_y)$, where 
$\rho^*$ is the complex conjugate of $\rho$.

The linear entropy of the sub-part $i$ $S(\rho_i)=1-{\rm Tr}\{\rho_i^2\}$ can 
be 
used as entanglement quantifier provided that the whole state is pure. It was 
put in relation with synchronization of chimera states 
in 
Ref. \cite{Viennot2016} in the case of a closed spin chain.

\subsection{Correlations of observables}\label{Sec_corr}

In Ref. \cite{Lee2013} the synchronization between coupled non-linear cavities 
($a$ and $b$) was addressed considering normalized intensity correlations 
$g_2(a,b)=\langle n_a n_b \rangle/(\langle n_a\rangle \langle n_b \rangle) $ 
between the cavities (first and second harmonic) modes. The average $\langle 
... \rangle$ was temporal in the classic limit (neglecting quantum noise and 
considering classical trajectories), and it was in this limit that 
synchronization was addressed. In the quantum regime, the expectation value 
over the quantum (steady) state was considered so that $g_2$ is a measure  of 
intensity correlations (capturing bunching/antibunching effects between the 
coupled systems). The transition between classical and quantum regime was 
described but addressing synchronization only in the classical regime and 
comparing it with steady state correlations when moving into the quantum regime. 

The average of the collective  operator 
\begin{equation}
 Z=\langle(\sigma_1^+\sigma_2^-+\sigma_2^+\sigma_1^-)\rangle
\end{equation}
was used in Ref. \cite{Zhu2015}  to detect  the presence of phase 
locking between two (ensembles of) spins and then the synchronization 
between them. There, it was shown that decay rates of these 
correlations encode information about the spectral content 
of the emitted radiation, which, in turn can be directly 
calculated using the two-time correlation function 
$Z=\langle(\sigma_1^+(\tau)\sigma_2^-(0)+\sigma_2^+(\tau)\sigma_1^-(0))\rangle$,
 as already done in Ref. \cite{Xu2014}. 

The value of spin-spin correlations
 $ \langle\sigma_1^\alpha\sigma_2^\alpha\rangle-\langle\sigma_1^\alpha\rangle\langle\sigma_2^\alpha\rangle$
 ($\alpha=x,y,z$) was also used by Hush \textit{et al.} \cite{Hush2015} as a sufficient 
criterion to assess the synchronization between the motion of two 
trapped ions. In such set-up, the spins represent the electronic 
degrees of freedom of the ions, and the value of their correlations 
was shown to be related to the relative phase distribution of the 
density matrix of the two motional degrees of freedom.

Looking at quantum correlations between observables to assess quantum 
synchronization is a natural strategy to identify nonclassical signatures but it does not always capture 
the emergence of similar time evolutions between different sub-systems. 
This approach is actually often considered looking at stationary states 
loosing any relation with the classical counterpart of synchronization.

\subsection{Kuramoto models}

The Kuramoto model \cite{Kuramoto1975} was introduced by Yoshiki Kuramoto to study the synchronous behavior of a large set of coupled oscillators, which appears naturally in the context of chemical/biological systems.
The equations of motion for the phase variable of the $N$ oscillators take the form:
\begin{equation}
 \dot{\theta}_i=\omega_i +\xi_i+\frac{K}{N}\sum_j \sin(\theta_i-\theta_j),
\end{equation}
where $\omega_i$ are oscillator frequencies, $\xi_i$ noise terms, and $K$ quantifies the coupling. An order parameter is defined assuming mean-field coupling
\begin{equation}\label{Rorder}
re^{i\psi}=\frac{1}{N}\sum_i e^{i\theta_i},
\end{equation}
where $r$ represents the phase-coherence of the population of oscillators. Moving to a rotating frame where $\psi=0$, the equations of motion become (without noise)
\begin{equation}
\dot{\theta}_i=\omega_i -K\, r\, \sin(\theta_i),
\end{equation}
which is just a particle in a washboard potential. This means that whenever $|\omega_i|<Kr$ the phase is trapped and we have synchronization, otherwise the phase slips down the washboard.
If the distribution of frequencies $g(\omega)$ is unimodal and centered around $\Omega$, the phase transition to complete synchronization occurs for the critical coupling value $K_c=2/\pi g(\Omega)$.

This model and similar ones have spurred a vast amount of research reviewed in \cite{Acebron2005}.
Here we note that whenever a system can be reduced to a set
of equations similar to the Kuramoto model, an analogous argument can be made to check whether there is phase locking or not. This is the case for coupled optomechanical oscillators 
as reported in Ref. \cite{Heinrich2011}.
The authors are able to reduce the mean field dynamics of the optomechanial array
with all-to-all couplings (after carefully eliminating the amplitudes from the dynamics) to a Kuramoto-type model, which for only two units simplifies to
\begin{equation}
\delta\dot{\theta}=-\delta\Omega-C\cos(\delta\theta)-K\sin(2\delta\theta),
\end{equation}
with $\delta\theta=\theta_2-\theta_1$ and $\delta\Omega$ the frequency difference. Once here, pure analysis of the parameters in the equation directly point to either phase locking or phase slip.

Notably, translation of the Kuramoto model to the semiclassical domain was recently achieved \cite{HermosodeMendoza2014} and it was shown that the picture of the washboard potential is a good intuitive guide, where now 
quantum tunneling can allow the phase to tunnel through maxima of the (otherwise phase-locking inducing) potential. This leads to the necessity of a higher critical coupling constant $K_c$ in order to 
pin down and lock the phases in the model.

\subsection{Other approaches}
\label{Sec_other}

Quantum synchronization has also been addressed considering phase space representations. In Ref.\cite{Ludwig2013}
the collective  phase-coherence among the components of an optomechanical array was characterized through  an order parameter, Eq.(\ref{Rorder}),
and looking at the transition of the Wigner representation from an angular-symmetric distribution to a coherent displaced state with time dependent phase.
Under the assumption of infinite non-linearity, Van der Pol oscillators in presence of driving were considered in  \cite{Lee2013a,Walter2015}
in strongly non-classical regimes,
where the dynamics is governed by few Fock states: the authors discuss the similarity of the Wigner distribution  
with   limit cycle appearing in the classical regime and the break of rotational invariance of $W$ to characterize phase locking. 
The marginals of the relative phase are also a signature of phase locking as considered in Ref.\cite{Hush2015}.
 Interestingly, phase state tomography has been recently reported in order to characterize 
phase diffusion and locking for  an on-fiber optomechanical cavity operating into the classical 
regime \cite{Shlomi2015}. Still, when writing this Chapter, there are not experimental  results 
on synchronization reported in the quantum regime.

\section{Synchronization of oscillators}

 \subsection{Linear networks of Brownian oscillators}
\label{SectLIN_HO}

Coupled harmonic oscillators dissipating in a bosonic environment represent 
 the simplest setup where synchronization of quantum systems can occur \cite{Benedetti2016}.
Their intrinsic linearity precludes the possibility of synchronous limit cycles, 
but not the presence of a long transient where synchronization is present
and also of steady oscillating states protected from dissipation. 
They have been studied for example in 
\cite{Giorgi2012,Manzano2013,Manzano2013a}, where it was shown that eigenmodes 
in the system of coupled oscillators can display
very different dissipation timescales for some parameter ranges. If one of the 
eigenmodes' dissipative rate is much smaller than that of the rest, this 
eigenmode dominates the dynamics of the coupled oscillators, and hence they 
become synchronized at the eigenfrequency of that mode. This synchronization
is temporary but can be very long if such rate is small \cite{Giorgi2012}. For certain situations 
of high symmetry, one of the eigenmodes of the coupled oscillators system
can be isolated from the environment and then synchronization can last forever \cite{Manzano2013}. 
Situations of higher symmetry can also occur where more than one eigenmode
is not dissipating, whereby synchronization can not happen \cite{Manzano2013a}. More elaborate situations 
have been studied, for example in harmonic networks \cite{Manzano2013} it 
has been shown that the full network or only a motif inside it can be 
synchronized by tuning one of the frequencies.

One of the main conclusions in this type of systems is that dissipation induces 
quantum synchronization through diffusive coupling:  be it for some 
parameter region or another, synchronization can always be achieved, with the sole
exception of the separate baths case, where each coupled oscillator is attached 
to its own independent heat bath. It is easy to demonstrate mathematically that
in this case, also eigenmodes are dissipating into equivalent independent heat 
baths, and therefore their dissipative rates are of equal size: no eigenmode
then
survives longer than the rest and thus synchronization is avoided.
Different dissipation mechanisms that can arise in extended environments
\cite{Galve2016} lead to specific forms of diffusive couplings that can induce synchronization.

The Hamiltonian that describes this dynamics is 
\begin{equation}
H=\sum_j \frac{p_j^2}{2m_j} +m_j\frac{\omega_j^2}{2}x_j^2 +\sum_{i\neq j} 
\lambda_{ij}(x_i-x_j)^2,
\end{equation}
whereas dissipation is introduced by coupling each oscillator to its own 
environment -eventually with different dissipation strengths-, to a common one -with a 
homogeneous coupling or not-, or other combinations. For ease of exposition here we 
will present the commonly denominated `common bath' case with system-bath interaction
\begin{equation}
H_{diss.}=\sum_j x_j \sum_k c_k Q_k,
\end{equation}
where all system components couple equally to the same  environment. In the case of
two units with equal mass and different frequency the system reads
\begin{equation}
H=\frac{p_1^2}{2m}+\frac{p_2^2}{2m} 
+\frac{\omega_1^2}{2}x_1^2+\frac{\omega_2^2}{2}x_2^2 +\lambda x_1x_2,
\end{equation}
where we have already absorbed quadratic contributions from the spring coupling 
into the oscillators frequencies.
The oscillators in the bath are uncoupled among them and have frequencies 
$\omega_j$, which together with the coupling constants $c_k$
define in the continuum limit what is usually called the spectral density 
$J(\omega)=\sum_k \frac{c_k^2}{\omega_k}\delta(\omega-\omega_k)$.
This function encodes basically all information about the dissipation 
characteristics and is usually taken to be `Ohmic' [that is, $J(\omega)\sim 
\omega$]
which gives a damping of the oscillators which is just proportional to their 
velocity. The linearity of the dynamics means that if we use an initial state 
which is Gaussian (all its information can be described from first and second 
moments), 
it will remain so at all times. Examples of Gaussian states  \cite{Ferraro2005,Adesso2007} widely used 
and relevant are displaced, squeezed and thermal states.

\begin{figure}[t!]
\includegraphics[width=0.7\textwidth]{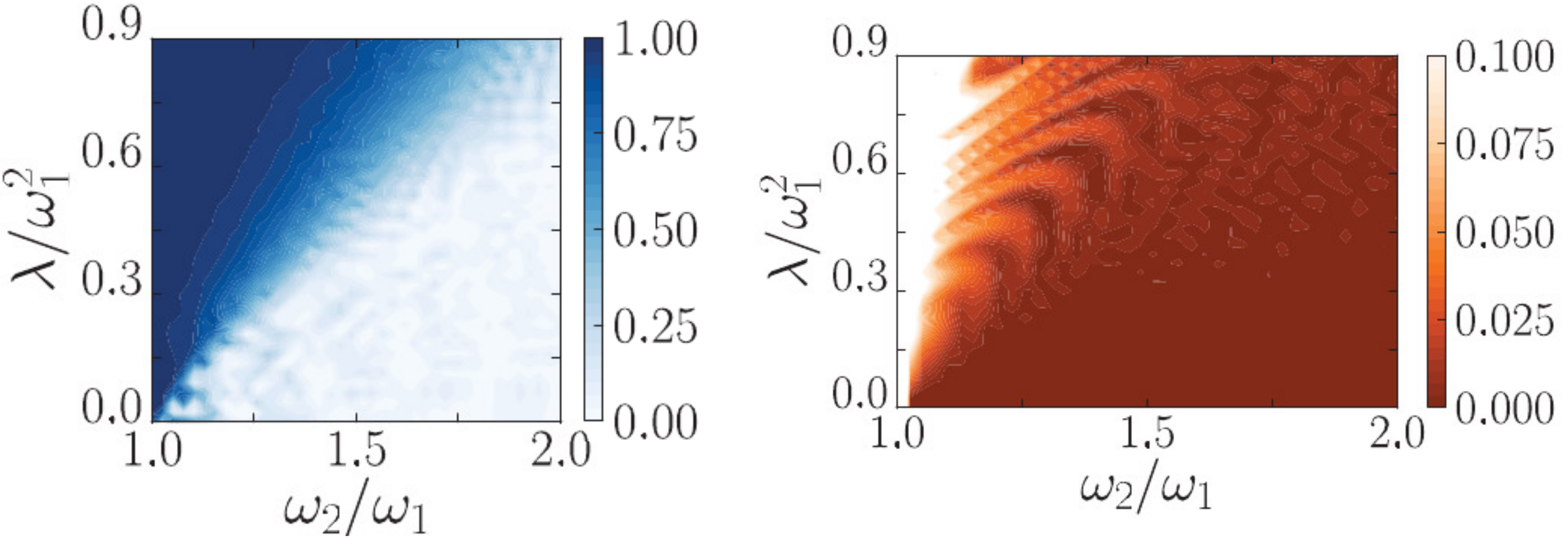}
\caption{Left: Pearson indicator of synchronization for different detunings and 
coupling strengths in the two coupled oscillators model dissipating into a 
common bath. They are 
initialized to squeezed vacuum states with zero first order moments. 
Synchronization is measured in the quadratures of each oscillators, i.e. 
$C_{\left<x_1^2(t)\right>\left<x_2^2(t)\right>}$ is plotted. Right: Quantum 
discord at the same value of time. A similar, although narrower, ``tongue" is  observed.}
\label{tongue}
\end{figure}

Synchronization can for example be displayed by the time evolution of second 
moments of quantum states, such as vacuum squeezed states, with the first moments 
being
zero at all times. An example of this can be seen in Fig.~\ref{tongue}, where 
the Pearson indicator is drawn for different detunings and coupling strength \cite{Giorgi2012}. 
The shape is
reminiscent of an Arnold tongue in classical physics, and has been also later 
observed with Van der Pol oscillators \cite{Ameri2015,Lee2013}. The fact that 
first moments are zero
while second moments are oscillating and synchronize does not seem to fit well 
with the  synchronization error indicator, which in fact grows in the regions
where the Pearson indicator clearly points to worse values. 

Furthermore, information-based correlations are preserved through the common 
eigenmode that does not dissipate, and thus survive longer in the synchronized 
case. 
However, correlations cannot assess synchronization if they were not already 
present at some initial time and thus are neither good synchronization measures 
for this example. We note that the considered system focuses on the effect of dissipative coupling in harmonic 
arrays, and this analysis could be applied  -for instance- the noisy precursor of an 
optomechanical system below the oscillation threshold.

\subsection{Self-sustained optomechanical oscillators}
\label{Sect_Self-sustainedOM}

An important example of synchronization dynamics are optomechanical systems: 
capable of displaying limit cycles of  constant amplitude, they provide 
an intuitive connection to what is known in the classical realm where the 
Kuramoto model is paradigmatic. Their ability to synchronize was first proven
in \cite{Heinrich2011}, where it was shown that mechanically coupled 
optomechanical oscillators above their dynamical Hopf bifurcation can be 
described 
with a Kuramoto-type model; phase and anti-phase synchronization are displayed. 
The analysis in the case of a common bosonic mode \cite{Holmes2012} and experimental demonstration of 
the classical synchronization of such system \cite{Zhang2012} followed shortly. 
Quantum synchronization has been later reported:
above a threshold mechanical coupling between 
optomechanical units, a regime of quantum phase-coherent
mechanical oscillations arises \cite{Ludwig2013}. The  measure of 
quantum synchronization analogue to synchronization error was considered for optomechanical 
oscillators in \cite{Mari2013}.

Optomechanical systems \cite{Aspelmeyer2014} comprise a mechanical mode and a confined optical mode 
which is typically driven by a laser field. These modes are
coupled nonlinearly through radiation pressure,  provided by a movable 
mirror or a structure which can be deformed by the action of light such as
a dielectric medium. The combination of an external pump and a nonlinear 
coupling provides stable limit cycles upon which synchronization can occur.
The usual form of the Hamiltonian describing such dynamics is
\begin{equation}
H=\Delta a^\dagger a +\omega b^\dagger b+ g a^\dagger a 
(b+b^\dagger)+iE(a-a^\dagger)
\end{equation}
with $a(a^\dagger)$ the annihilation (creation) operators of light in the 
cavity, $b(b^\dagger)$ those of the mechanical mode and $E$ is the intensity of 
the laser
input. The Hamiltonian is written in the frame rotating with the laser frequency 
$\omega_L$, which is detuned by $\Delta=\omega_C-\omega_L$ with respect to the 
cavity frequency $\omega_C$. Both modes are further coupled to noise sources 
with strength $\kappa, \gamma$ respectively, providing dissipation.
Below a given threshold intensity of the laser, the amplitude of light and 
mechanical modes simply decay to the value enforced by the noise sources. 
However, above
that intensity, multistability and limit cycles of increasing complexity can be observed. The final 
ingredient to observe synchronization is coupling, via the mechanical or optical 
degrees of freedom,
several optomechanical systems. For illustrative purposes we choose here 
mechanical coupling of the type $H_{int}=\mu (b_1b_2^\dagger+b_1^\dagger b_2)$.
We will follow here the example of dynamics given in \cite{Mari2013} to compare 
different measures of synchronization and correlations. After a transient, limit 
cycles are achieved for each
degree of freedom and we can linearize them around their stable orbits with the 
usual ansatz: $\hat{a}_j=A_j(t)+\delta\hat{a}_j$ and 
$\hat{b}_j=B_j(t)+\delta\hat{b}_j$, with capital letters
representing the limit cycles as a classical variable, and $\delta\cdot$ the 
linear displacements (fluctuations) with respect to them. The ansatz is then 
used to neglect nonlinear terms in the dynamics.
Finally, the fluctuations can be arranged in the form of a covariance matrix 
whose time evolution can be integrated, yielding both the dynamical content and 
the quantum/classical information content.

\begin{figure}[h!]
 \includegraphics[width=\textwidth]{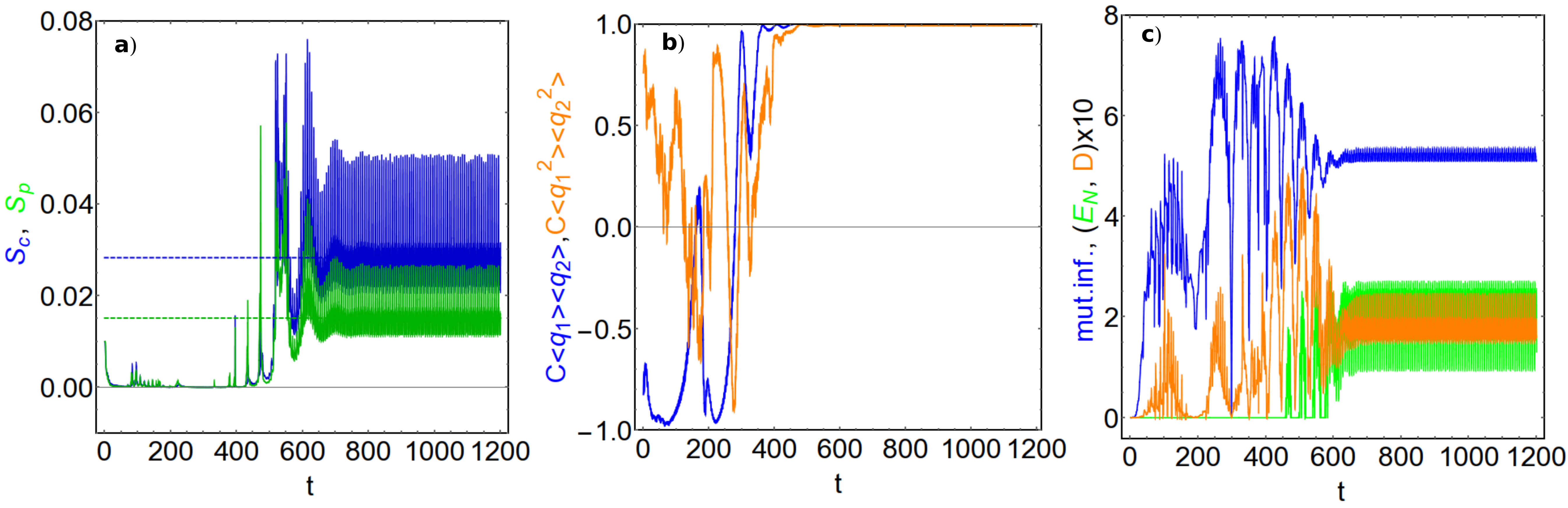}
 \caption{Comparison of several synchronization measures: (left) error 
synchronization $S_c$, $S_p$, (middle) Pearson indicator for first and second 
momenta 
 $C_{\left<q_1\right>\left<q_2\right>}$, 
$C_{\left<q_1^2\right>\left<q_2^2\right>}$,
 and (right) correlations: mutual information, logarithmic negativity and 
quantum discord. We set parameters $\omega_1=1$, $\omega_2=1.005$, 
$g=0.005$, $\mu=0.02$, detunings $\Delta_j=\omega_j$, 
 $\kappa=0.15$, $\gamma=0.005$ and laser input $E=320\kappa=48$, as in Ref. 
\cite{Mari2013}. The initial condition for first moments are 
 $\left<q_1(0)\right>=100$ and $\left<q_2(0)\right>=-100$, and all other first 
moments zero. The 
 second order moments at $t=0$ are 100 times their vacuum value. Changing these 
initial conditions do not change qualitatively the results.}
 \label{figOPTOMECH}
\end{figure}

Coupled optomechanical oscillators represent a good platform 
combining sufficiently complex dynamics and the possibility to assess different
informational measures, so that 
proposed indicators of synchronization can be compared, as seen in 
Fig.~\ref{figOPTOMECH}. We consider two optomechanical systems like in
Ref. \cite{Mari2013}. 
The qualitative behaviour is similar for all 
indicators: an initial build up and a final stage of full 
synchronization by a stationary value of all quantities. 

\begin{figure}[t!]
 \includegraphics[width=\textwidth]{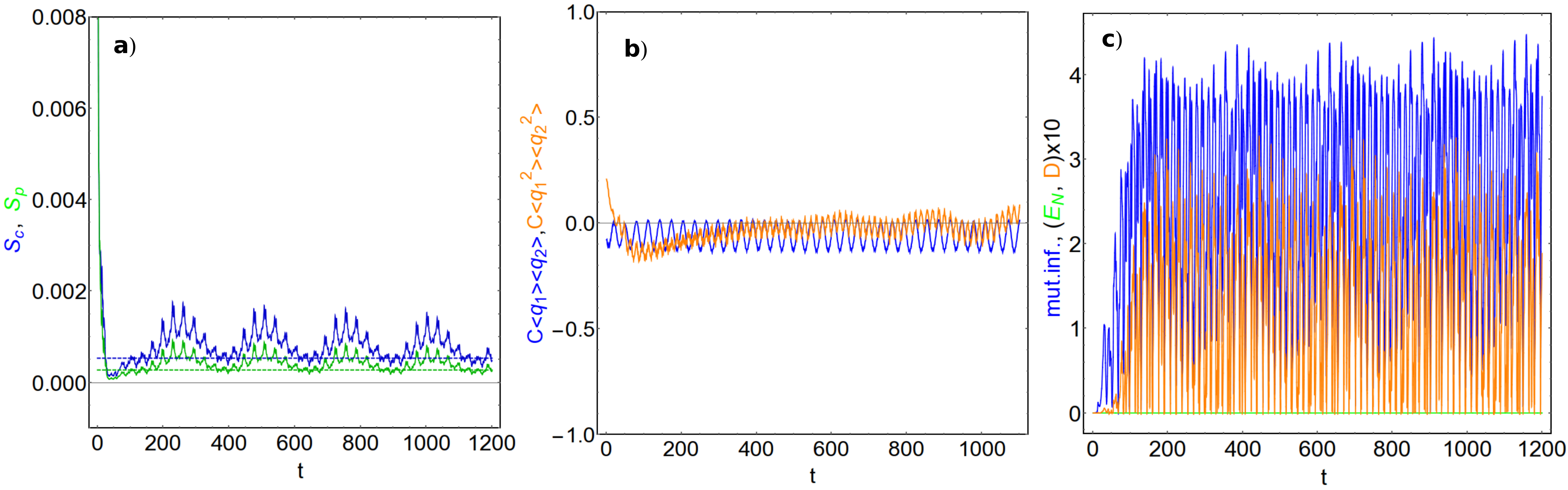}
 \caption{Same parameters as Fig.~\ref{figOPTOMECH}, but with higher detuning 
$\omega_2=1.2$, so there is no synchronization. Notice that the scale in (a) is ten-fold lower than in Fig.~\ref{figOPTOMECH}a.}
 \label{figOPTOMECH2}
\end{figure}

As mentioned before, perfect synchronization can be 
identified looking at the absolute value of the indicator only for the Pearson coefficient 
(Sect. \ref{Sec_Pearson}), which really 
shows perfect synchronization (value $\sim 1$) for first and second moments of the mechanical
observables (Fig.~\ref{figOPTOMECH}b, time $\gtrsim 500$). 
The synchronization error indicators have a rather low value, Fig.~\ref{figOPTOMECH}a,
compared to their maximum attainable  1 (see Sect. \ref{Sec_error}), and we have checked that for different 
initial conditions of the 
mechanical first moments, it can be increased, with very similar qualitative 
behaviour. The possibility to reach its maximum bound Eq. (\ref{upper}) in such system
has not been reported, although for fixed initial conditions it might
be informative to assess the quality of synchronization
when changing Hamiltonian parameters (as when comparing 
Fig.~\ref{figOPTOMECH} and Fig.~\ref{figOPTOMECH2}).
As a note, while these different indicators can signal stable 
synchronization for $t>600$, the Pearson coefficient spot it sooner.
Comparing figures ~\ref{figOPTOMECH}c and  \ref{figOPTOMECH2}c
we can conclude too that any measure based on 
classical/quantum correlations (of information-theoretic character) is not of 
too much value if we compare absolute
values. Introducing initial squeezing in the mechanical modes changes the time 
profile of all correlations, but as expected not their final stable values.

Finally, what is important to note  is that the stability in time of the indicators is a necessary 
condition for signaling synchronization, and once that is achieved, 
comparison of absolute values
might yield some extra information, although only the Pearson indicator 
is a bona fide absolute measure in this sense.

\section{Synchronization of interacting spins}\label{spinsec}

When considering precessing spins, the behavior of local spectra can be used to extract direct information about 
the presence of a synchronized dynamics. This method was adopted by Orth and 
coworkers in Ref. \cite{Orth2010},  where they considered two interacting spins 
dissipating through a common environment, and observed that there is a regime in 
the parameter space of the system where only a single frequency  appears in the 
spectrum of both the local observables. This single-line spectrum is not the 
only possible manifestation of synchronization. It is indeed possible, like in 
the infinite-dimension Hilbert space cases discussed above, that the collective 
dynamics favours the suppression of some spectral lines, while one of them has a 
very long life-time, leaving the system synchronized during the transient decay 
leading to steady state \cite{Giorgi2013,Giorgi2016}.  In these cases, the 
Pearson's measure can be adopted to observe the dynamical setting-up of 
synchronization. It can be defined considering the expectation values of
any arbitrary operator for each spin 
$A_
k$, $k=1,2$, decomposed in the single-spin basis $\{\sigma_x^k, \sigma_y^k, 
\sigma_z^k, I^k_d\}$:
\begin{equation}\label{OperatoreArbitrario}
  A_k= a_x^k \sigma_x^k + a_y^k \sigma_y^k + a_z^k \sigma_z^k + a_d I^k_d.
\end{equation}
Actually, in many cases, it is enough to consider one spin direction. For 
instance, in Ref. \cite{Giorgi2013}, the $z$ components of the two spins were 
synchronized from the beginning and the interesting part of the dynamics 
concerned the $x-y$ plane. 

In the following, we are going to compare the Pearson's measure of quantum 
synchronization with correlation indicators in a model of two detuned  spins 
interacting through an Ising-like coupling: 
\begin{equation}
H_S=\frac{\omega_1}{2}\sigma_1^z+\frac{\omega_2}{2}\sigma_2^z+\lambda \sigma_1^x 
\sigma_2^x. \label{htot}
\end{equation} 
Let us assume that the spins experience a dissipative dynamics induced by the 
presence of a thermal environment weakly coupled to the system through
\begin{equation}
H_I=  \sum_k g_k (a_k^\dag+ a_k )(A\sigma_1^x +\sigma_2^x).
\end{equation}
Here, the annihilation (creation) operators $a_k$ ($a_k^\dag$) act on the bath 
degrees of freedom and the coefficient $A$ determines the the ratio between the 
strength of the two spin-bath couplings.  For the sake of clarity, in the 
following we will only discuss the two extreme cases $A=0$ (local environment) 
and $A=1$ (common environment). 
In order to derive a master equation describing the dynamics of the  two spins, 
it is necessary to know the diagonal form of $H_S$, which can be obtained 
applying the standard Jordan-Wigner transformation, 
mapping spins into spinless fermions, defined as $\sigma_1^z=1-2 c_1^\dag c_1,\, 
\sigma_2^z=1-2 c_2^\dag c_2,\,\sigma_1^x= c_1^\dag+ c_1,\,\sigma_2^x= (1-2 
c_1^\dag c_1) (c_2^\dag+ c_2)$ \cite{Lieb1961}. We have
\begin{equation}
H_S=E_1 (\eta_1^\dag \eta_1-1/2)+E_2 (\eta_2^\dag \eta_2-1/2),
\end{equation}
with
$E_1=\frac{1}{2}\left(\sqrt{4\lambda^2+\omega_+^2}+\sqrt{4\lambda^2+\omega_-^2}
\right)$ and $E_2=\frac{1}{2}\left(\sqrt{4\lambda^2+\omega_+^2}-\sqrt{4\lambda^2+\omega_-^2}
\right)$
where $\omega_\pm=\omega_1\pm \omega_2$. The quasi-particle fermion operators 
are obtained combining  the Bogoliubov transformation $c_1=\cos \theta_+ \xi_1+ 
\sin\theta_+ \xi_2^\dag,\,c_2=\cos \theta_+ \xi_2- \sin\theta_+ \xi_1^\dag $
together with the rotation 
$\xi_1=\cos\theta_-\eta_1^\dag+\sin\theta_-\eta_2^\dag,\,
\xi_2=\cos\theta_-\eta_2^\dag-\sin\theta_-\eta_1^\dag$.

The spectral density  $J(\omega)= \sum_k g^2_k\delta(\omega-\Omega_k)$ is 
assumed to follow, apart from a high-frequency cut-off, the Ohmic power law 
$J(\omega)\sim \omega$.  Assuming weak dissipation, the  qubit pair  dynamics 
can be  studied in the Born-Markov 
and secular approximations \cite{Breuer2007}
with Lindblad master equation $\dot\rho(t)=-i [H_{S}+H_{LS},\rho(t) ]+{\cal 
D}[\rho(t)]$, where the
Lamb shift  $H_{LS}$ commutes with $H_{S}$
and  where ${\cal D}[\rho(t)]$, which takes into account dissipation, is the 
sum of four terms, each of them associated to one of the four transition 
frequencies $\pm E_i$ $(i=1,2)$:
\begin{equation}
{\cal D}(\rho)=\sum_{i=1}^2\tilde\gamma_i^+ {\cal 
L}[\eta_i](\rho)+\sum_{i=1}^2\tilde\gamma_i^- {\cal L}[\eta_i^\dag](\rho),
\end{equation} 
Here,  the Lindblad superoperators are defined as 
$ {\cal L}[\hat X](\rho)=\hat X \rho \hat X ^\dagger -\{\hat \rho ,\hat 
X^\dagger \hat X \}/2$.
 The exact value of the decay rates $\gamma_i^\pm$ will depend on the nature of 
the system-bath coupling. In the case of local dissipation $A=0$, their specific 
form can be found in Ref. \cite{Giorgi2016}. As already discussed, 
synchronization takes place if there is substantial separation between the two 
largest 
$\gamma$'s determining the dynamics. In this case, local degrees of freedom 
undergo quasi-monochromatic oscillations and  their relative phases  get 
locked.

\subsection{Spin synchronization}

As discussed in Ref. \cite{Giorgi2016}, the case of a local environment $A=0$ 
shows an ``anomalous" synchronization pattern. Indeed, unlike classical 
manifestations of synchronization and quantum synchronization induced by a 
common environment (Refs. \cite{Giorgi2012,Manzano2013,Giorgi2013}), it is 
greatly enhanced in the strong detuning $\Delta=|\omega_1-\omega_2|$ regime, 
while the direct spin-spin coupling $\lambda$ has a partially detrimental 
effect. Turning attention to the case $A=1$, the presence of a common bath has 
the tendency to facilitate spontaneous synchronization. In order to observe it, it is fundamental for the 
two spins to  have an interaction strong enough as to compensate the detuning. 
Furthermore, in the local-bath case, depending on the Hamiltonian parameters, 
synchronization can appear both in phase and in anti-phase. This feature is 
suppressed in the presence of a common environment, where only 
anti-synchronization can be observed. This is due to the different interplay 
between the $\gamma$'s  in the two 
scenarios.

\begin{figure}[t]
\includegraphics[width=14cm]{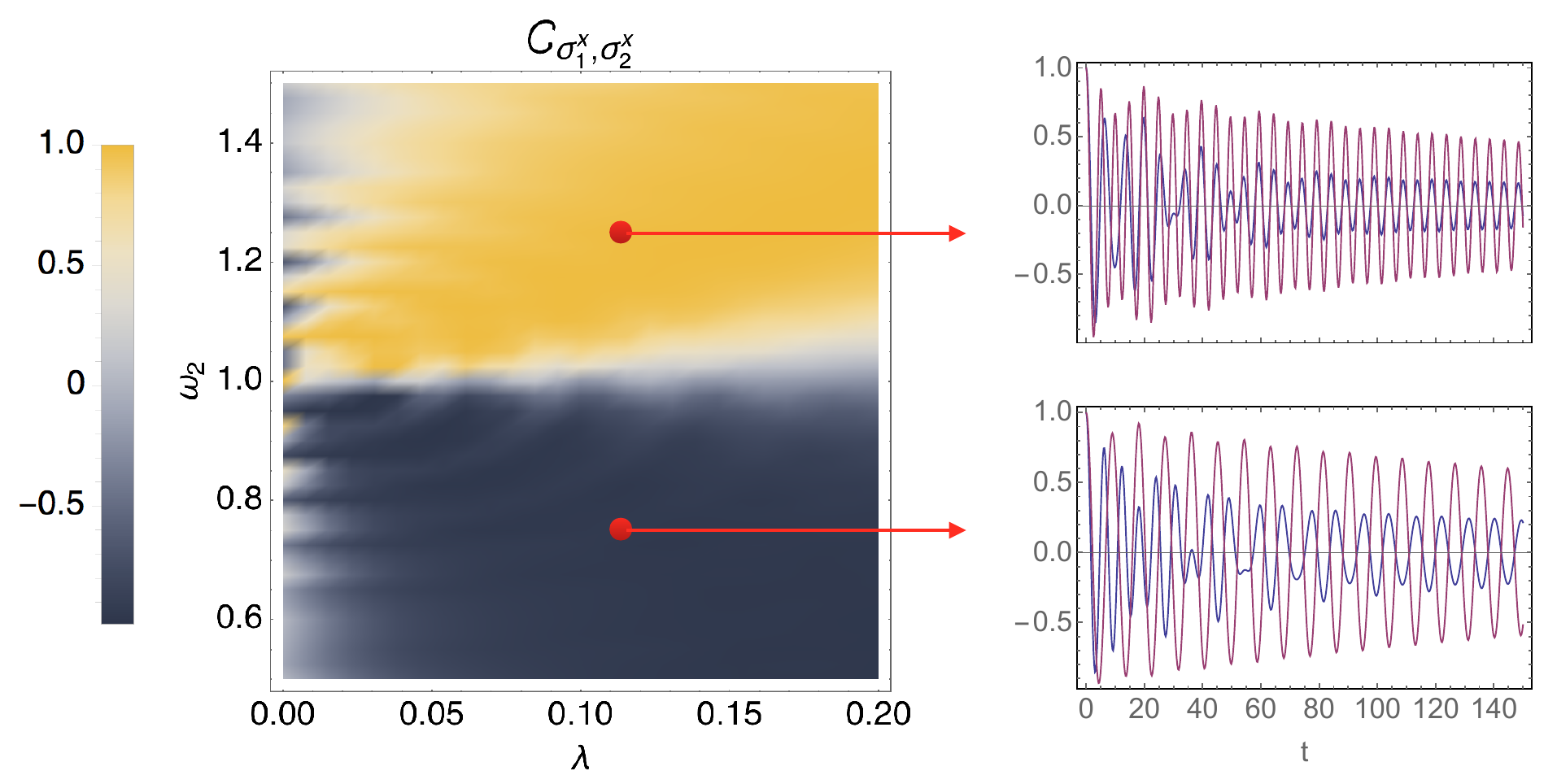}
\caption{Left panel: synchronization diagram as a function of $\omega_2$ and 
$\lambda$ for a local bath. ${\cal C}_{\sigma_1^x,\sigma_2^x}$ has been 
calculated at time $t=75$ (in units of $\omega_1$) using a time window of 
$\tau=10$. As explained in the text, the local bath ($A=0$) is assumed to be Ohmic with 
cut-off frequency $\omega_c=20$, and its temperature is $T=0$. The initial state 
is 
$|\psi(0)\rangle=(\ket{\uparrow}+\ket{\downarrow})(\ket{\uparrow}+\ket{
\downarrow})/2$. Left panels: $\sigma_1^x(t)$ (blue) and  $\sigma_2^x(t)$ (red) 
assuming  $\omega_2=1.25$ and $\lambda=0.11$ (top) and $\omega_2=0.75$ and 
$\lambda=0.11$ (bottom).}
\label{figspin1}
\end{figure}

The anomalous synchronization emerging from a local environment is shown in the 
$\lbrace \Delta-\lambda\rbrace$ diagram of Fig. \ref{figspin1}. The local 
observable used to calculate the Pearson's parameter ${\cal C}$ are, 
respectively, $\sigma_1^x$ and  $\sigma_2^x$. even if the calculation could be 
extended to generic local operators without qualitative changes in the results.
In the left panel, we show  the synchronization diagram, showing the transition 
from phase to anti-phase, while in the right panels we plot the trajectories of 
the two local observables in the two distinct regimes.

\begin{figure}[t]
\includegraphics[width=8cm]{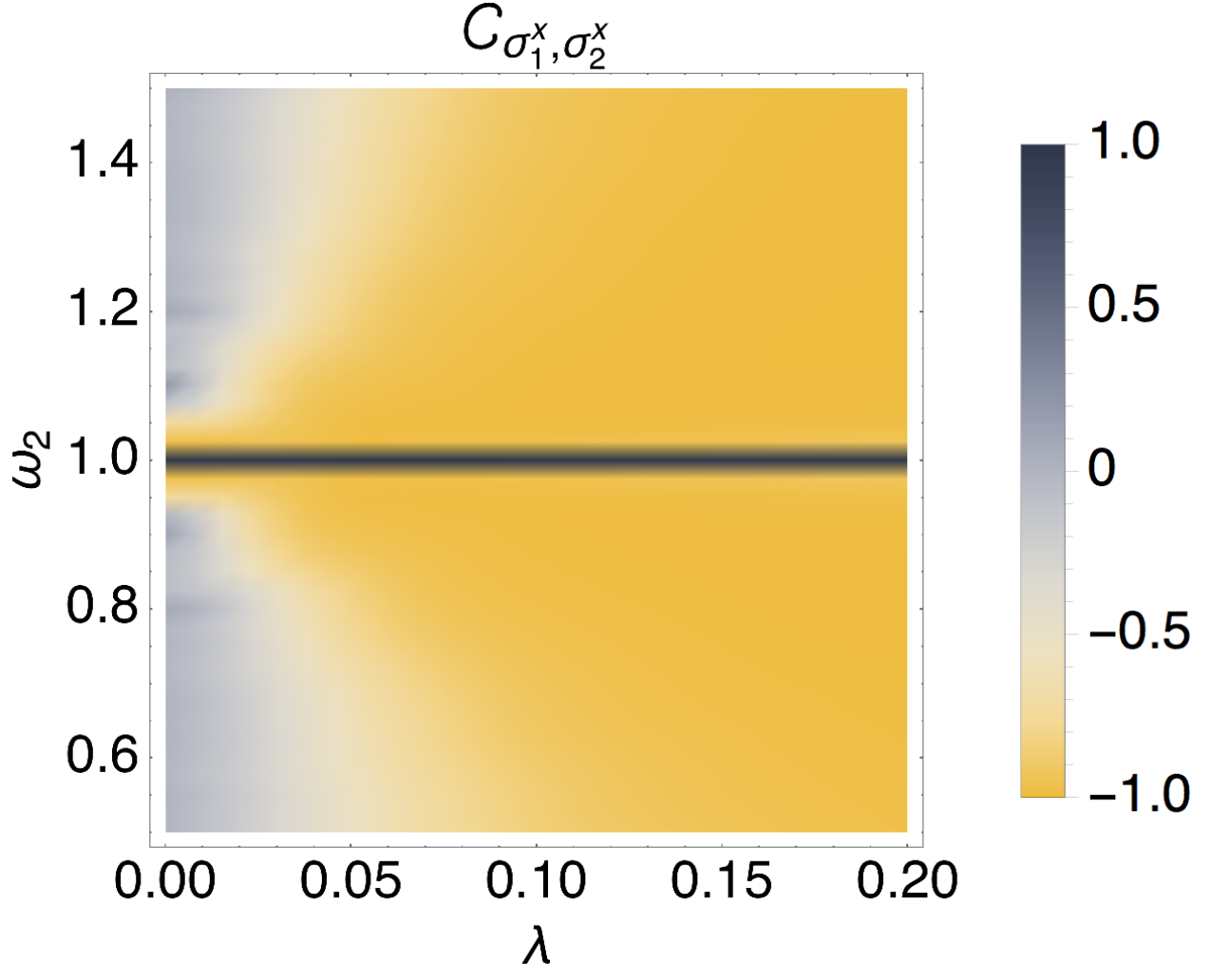}
\caption{${\cal C}_{\sigma_1^x,\sigma_2^x}(t=75,\tau=10)$ as a function of 
$\omega_2$ and $\lambda$ in the presence of a common bath ($A=1$). All other 
conditions as in  Fig. \ref{figspin1}.}
\label{figspin2}
\end{figure}

 On the other hand, the behavior of ${\cal C}$ for a common bath is displayed in 
Fig. \ref{figspin2} and it is very much similar to the characteristic Arnold 
tongues emerging in classical synchronization problems. In this case,  
anti-synchronization emerges provided that the spin-spin coupling is not too 
small with respect to the detuning. As a singular behavior, around $\Delta=0$, 
the system shows "trivial" synchronization, given that the two spins become 
indistinguishable.

The complementarity and the qualitative difference of the synchronization 
diagrams emerging in the two cases under study  will be used in the following of 
this section to compare the Pearson's measure to correlation quantifier, namely, 
spin-spin correlations $\langle \sigma_i^+\sigma_j^-\rangle$, mutual 
information, and entanglement.

\subsection{Spin correlations}

As noticed in Ref.  \cite{Zhu2015}, 
$Z=\langle\sigma_1^+\sigma_2^-+\sigma_2^+\sigma_1^-\rangle$ plays the role of a 
phase locking indicator when applied to interacting spins,  and then can be used 
as a synchronization measure (see Sec. \ref{Sec_corr}). At a first sight, ${\cal C}$ and $Z$ evolve 
independently, as the sets of equations of motion of their respective matrix 
elements are not coupled  to each other. Actually, the constraints, to which a 
physical density matrix is subject to, make the behavior of the two indicators 
very close to each other. This aspect is discussed in great detail in 
Ref.  \cite{Bellomo2016}, in the case of independent spins,  where the interplay 
between spontaneous synchronization  and superradiance is studied.
In fact, in order for  spontaneous synchronization to emerge,the whole system needs to support a 
long-lasting collective mode, which unavoidably displays spin-spin correlations.

\begin{figure}[t]
\includegraphics[width=9cm]{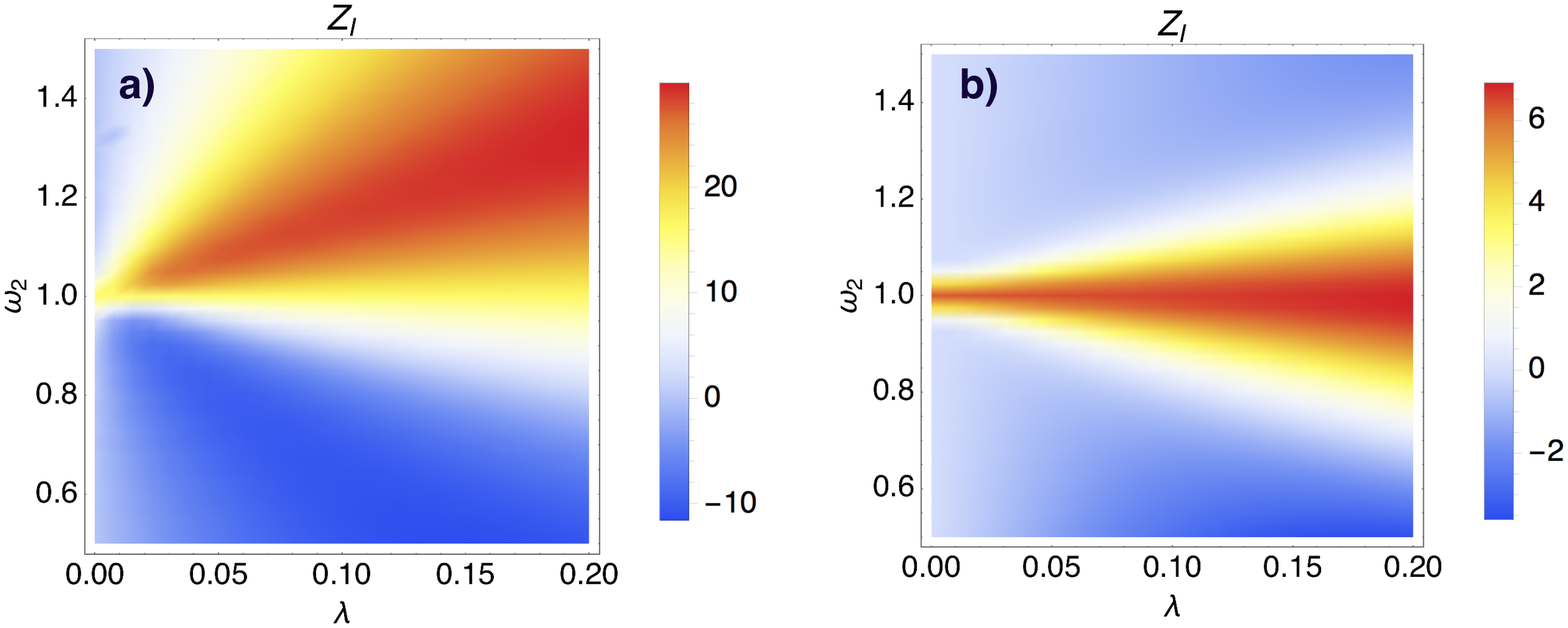}%
\includegraphics[width=9cm]{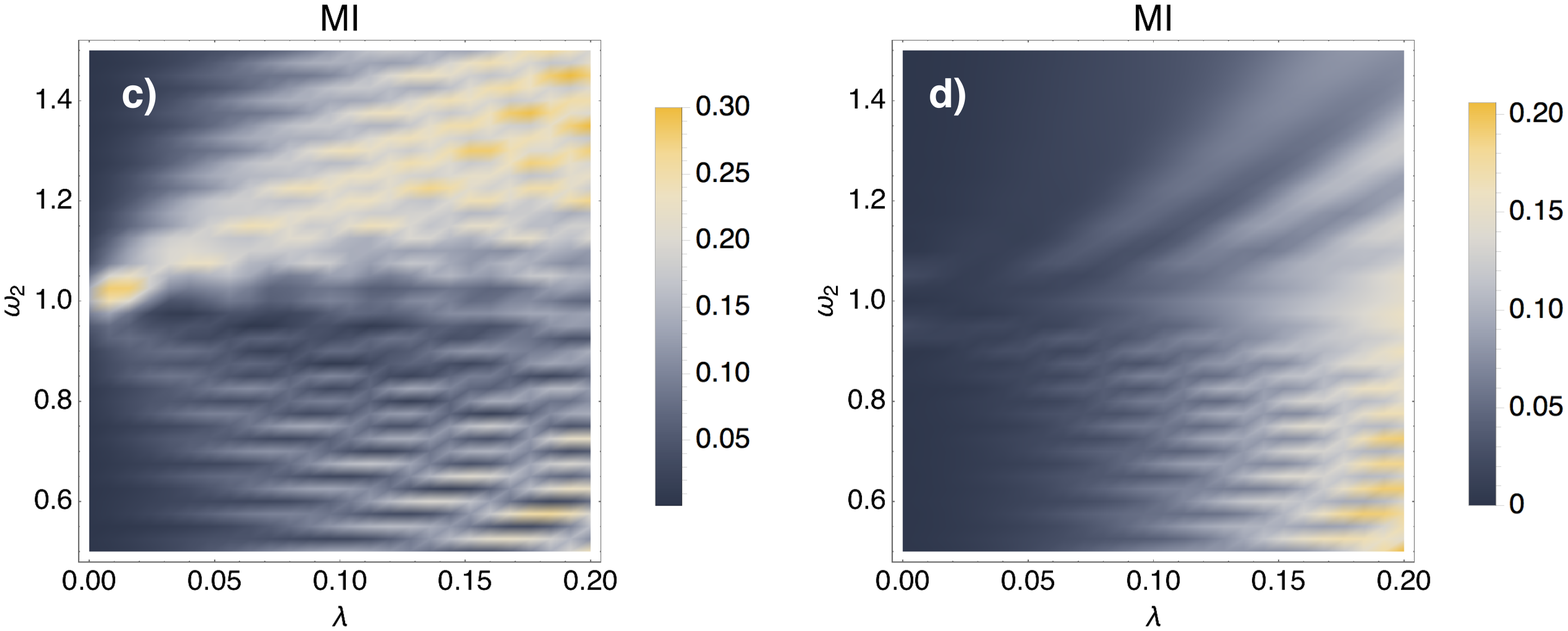}
\caption{$Z_I$ (a,b) and mutual information $MI$ (c,d) as a function of $\omega_2$ and $\lambda$  for a local bath $A=0$
(a,c) and for a common bath $A=1$ (b,d). All parameters as in Fig. \ref{figspin1}.}
\label{figspincorr}
\end{figure}

These qualitative considerations are confirmed in the cases we are discussing 
here: in both models (of local and global dissipation), low-quality 
synchronization  ${\cal C}$ is always accompanied by a 
value for $Z$ close to zero. On the other hand, within the synchronized regions, 
  $Z$  is significantly enhanced. Furthermore, the sign of $Z$ is reminiscent of 
the phase--anti-phase form of synchronization. These results are shown in Fig. 
\ref{figspincorr}, where, in order to wash out faster oscillations and 
regularize the picture, the time integral of $Z$ ($Z_I=\int_{t=0}^{100} 
Z(t^\prime)dt^\prime$)  is plotted as a function of  $\omega_2$ and $\lambda$. 
Both cases of local ($A=0$) and global ($A=1$) dissipation are shown. For a local bath, Fig. 
\ref{figspincorr}a,
the change from positive to negative values for the spin-spin correlation 
parameter takes place in the same region where ${\cal C}$  passes from synchronization to 
anti-synchronization (compare with Fig.  \ref{figspin1} a,b). In view of 
the previous 
considerations, this change is suppressed for $A=1$, Fig. 
\ref{figspincorr}b. It is   worth 
noticing that the "anomalous" synchronization peak around $\omega_1=\omega_2$ dispayed by   ${\cal C}$ (Fig. \ref{figspin2}) is 
broadened by $Z_I$ (Fig. 
\ref{figspincorr}b), whose value appears smoother at critical changes.

\subsection{Mutual information and entanglement}

Mutual information ($MI$) as a quantum synchronization witness in spin systems was 
proposed by Ameri \textit{et al.} in a work dealing with a system of two qubits 
placed in two coupled  cavities where only the first one is driven by a laser, 
while the second one is populated by the photons leaking
from the first cavity \cite{Ameri2015}.  In that example, it was shown that the 
steady-state mutual information was reminiscent of the synchronized oscillations 
of local operators in the pre-steady-state regime.

In the following we show that, in the models we are investigating, $MI$ does not 
play the witnessing role suggested in Ref. \cite{Ameri2015}. As a first 
observation, we notice that we deal with systems decaying towards an equilibrium 
state (the Gibbs state) that only depends on the Hamiltonian parameters, while 
the synchronization diagram depends critically on the properties of the 
environment. To make this point clearer, both cases we are considering here 
admit the same equilibrium state, whilst the two  synchronization diagrams are 
radically different. One may ask if some information about synchronization 
appears in the dynamical behavior of $MI$ instead of its asymptotic value.  For 
this reason, we considered the time at which synchronization starts to be solid 
($t=80$ in Fig. \ref{figspincorr} c,d) and calculated $MI$ for the two models (at longer times $MI$ would rapidly 
converge to zero everywhere). The two behaviors  indicate a very weak connection 
between $MI$ and  spontaneous synchronization. Starting from a factorized state, the coupling $\lambda$ 
immediately 
produces a quite robust amount of $MI$ (depending on the strength of $\lambda$) 
between the two spins. Then, the presence of dissipation makes this correlation 
disappear, but it seems  that the way $MI$ fades away is not connected with  the 
building up of a synchronized dynamics. 
A very similar argument can be applied to entanglement, that can be quantified 
using the concurrence $E$ \cite{Wootters1998}.  The dynamical behavior of $E$ 
(not shown) displays qualitative features very close to the ones on $MI$.

Besides the specific models studied here, the argument can
be made more general considering the case of a purely dephasing dynamics. On the
one hand, as discussed in Ref. \cite{Giorgi2013}, such a process is not able to
induce any synchronization, as no time scale separation takes place. On the
other hand, entanglement and mutual information can converge to a finite value
under the same circumstances.
Therefore quantum correlations in the
steady state, in general, do not witness a
previous syncronization during relaxation.

\section{Discussion and conclusions}
Present research on quantum synchronization has just started to unveil
the distinctive features of this phenomenon. Several factors, known to influence
it in the classical regime \cite{Pikovsky2001}
are under study, including, for instance, non-linearity, dissipation,
noise, forcing, mutual or directional coupling between inhomogeneous components,
 or time delay. Experiments reporting distinctive signatures of quantum
synchronization are expected to flourish in the next years.

The question about
what is essential of quantum synchronization with respect to the classical one
is intimately related to the interplay between temporal and quantum
correlations. From the previous analysis we can establish few relevant criteria
to approach the phenomenon of quantum synchronization as described by different measures
and to assess usefulness and meaningfulness in each specific context:

\begin{itemize}
 \item{\bf Absolute reference value:} In order to be able to assess the amount of
 synchronization in different regimes, it is important for a measure to be bounded
 and to have a definite value associated  to  the perfect emergence of full
 synchronization. The Pearson's parameter Eq. (\ref{eq:pears})
 reaches values very close to the maximum attainable  $ |{\cal C}_{A_1,A_2}(t|\Delta t)|\simeq 1$ whenever good
synchronization emerges. Similarly, the synchronization error  Eq. (\ref{syncerr}) is bounded in the quantum case,
whereas the classical one is not. Still, reported values are rather modest
 in the case of coupled optomechanical oscillators \cite{Mari2013} and saturation of this bound that would
correspond to the best quantum synchronization has not yet been reported.

 \item{\bf Time dynamics dependence:} The concept of synchronization is relative 
 to the time evolution of system's observable or variables
 and a measure of synchronization should reflect it covering a time window of the system dynamics, like in temporal averages for instance, 
or being robust during evolution.
 This is the case for several measures in different ways: some are based on  time averages 
(e.g.  Pearson's parameter (\ref{eq:pears})), others maintain distinct
 higher values during synchronization  (e.g. synchronization error (\ref{syncerr})), and others assess the time stability
of the process (as Lyapunov exponents \cite{Li2015}).  In general, the problem when looking at instantaneous (quantum) correlations, $MI$ etc.
is that they can be instantaneously
huge even when there is no synchronization, as shown in Fig. \ref{figOPTOMECH2}.
On the other hand, looking at asymptotic values is not always leading to an insightful synchronization condition. 

\item{\bf Local vs. non-local} Among the reported measures of synchronization, some refer to
local observables
 of the synchronized systems (like Pearson factors, or local phases in Kuramoto models) while other
 refer to quantum correlations present in the composed system (in this sense being 'non-local').
 The possibility to associate a genuine quantum correlation to synchronization is clearly appealing
 to distinguish it from classical synchronization, as for instance with the synchronization error \cite{Mari2013}.
 On the other hand, this can give rise to spurious definitions of quantum synchronization, actually not related at all 
with this dynamical phenomenon. This question is still open and few further considerations are given below.
\end{itemize}

In the attempt to identify a measure for a genuine quantum synchronization,
different indicators have been proposed that actually do not refer to observables
but to the full quantum state, as discussed in Sec. \ref{sec3}.
Invoking generic quantum correlation as a measure of synchronization is in general
not convincing. As an example,
a bipartite Bell-state is strongly correlated under any possible definition of
correlation, but in general this has nothing to do with synchronization.
As a matter of fact,  any local unitary would leave it unchanged, while altering
the dynamics of the components of the system can alter
 dynamical synchronization.
Even if non-local correlations are not necessarily associated to specific
synchronization phenomena,
it is important to remind that quantum correlations and dynamical
synchronization can occur under the same conditions in some systems
\cite{Giorgi2012,Manzano2013,Mari2013,Lee2014}.
Still, quantum correlations that capture specific signatures of synchronization are not  
generally established.
Looking at the examples we have treated  here we can also draw some conclusions.

Optomechanical self-sustained oscillators can achieve synchronization and several parameters like
the error and Pearson indicators give a similar insight to mutual information or other 
correlations (Fig. \ref{figOPTOMECH} and \ref{figOPTOMECH2}). Synchronization error however dispays a small  value far for the maximum bound 
not providing an absolute indication for the synchronization degree (Fig. \ref{figOPTOMECH}a). Furthermore, correlations signal synchronization 
not by their value, but by having a final stable nonzero value, in contrast to a highly oscillating one
in the case of no synchronization.

In the case of linear dissipative oscillators the situation worsens. The Pearson factor gives
a valuable guide to look for synchronization, while all the other indicators fail: synchronization error
does not provide a good estimate of the behaviour of synchronization with respect to the system's parameters
and other information measures are strongly initial state dependent. Still, robust quantum correlations (such as discord) 
can witness synchronization, as both emerge under the same circumstances in this case.
At some level, all of the measures can be used to yield some insight, however they require some craftmanship efforts
as compared to Pearson coefficients.

The literature about quantum synchronization in spins is much more limited with
respect to the case of harmonic oscillators or optomechanical systems.
Furthermore, in many cases, synchronization has been assessed using ad-hoc witness
measures more than quantifiers. This chapeter represents the first attempt to compare such
quantities. As a result, we observed consistent indication of synchronization between Pearson's and  spin-spin $Z$ 
indicators, due to a strong interplay between phase-locking dynamics and the dynamics of the local observables. 
In contrast, mutual information and entanglement fail to give any useful information. 

Finally, all the previous measures of synchronization could be modified to account for
more general forms of synchronization.
In all the discussed cases, synchronization is either in-phase or anti-phase.
It is worth remarking that, in general, delayed synchronization can also emerge
and synchronization  indicators need to be improved  to catch this effect. This can be easily done,
for instance in the case of the Pearson's parameter, allowing one of the two sliding windows
to open at a time different from the other one, that is equivalent to delay the time
of one of the observable expectation value 
\begin{equation}
 {\cal C}_{A_1 (t),A_2(t+\tau)}(t|\Delta t).
\end{equation}
Similarly this could be done for all indicators based on local observables.
Another way of improving synchronization indicators
consists in correcting possible relative amplitude mismatch effects,
similarly to the conditional variance factor appearing in the context of the EPR correlations
\cite{Drummond1990}.

To conclude we would like to stress that the field is still in its early stages
and this work is the first attempt to assess meaning and utility of different synchronization measures
as well as their possible dependence to 
the specific features of the system under study. Up to now, no experimental results in the quantum domain are at hand. Therefore 
there is plenty of room for improvement and surprises, both regarding the theoretical framework and
possible practical applications of potential use as quantum technologies.

Funding from EU project QuProCS (Grant Agreement No. 641277), MINECO (Grant
No. FIS2014-60343-P and FEDER project QuStruct FIS2015-66860-P), and 
``Vicerectorat d'Investigaci\'o  i Postgrau"  of the UIB are acknowledged.

%

\end{document}